\documentclass[pra, 11pt, eqsecnum,floatfix, showpacs, showkeys, nofootinbib]{revtex4}
\usepackage{graphics}
\usepackage{bm}

\newcommand{\gest}{g_{\rm est}}
\newcommand{\numin}{\nu_{\rm min}}

\begin{document}

\title{Qubit metrology and decoherence}
\author{Anil Shaji}
\affiliation{Department of Physics and Astronomy, University of New Mexico, Albuquerque, NM 87131}
\author{Carlton M. Caves}
\affiliation{Department of Physics and Astronomy, University of New Mexico, Albuquerque, NM 87131}
\date{\today}

\pacs{03.65.Ta, 06.20.-f, 03.65.Yz, 03.67.-a}
\keywords{quantum metrology, decoherence, Braunstein-Caves inequality}

\begin{abstract}
Quantum properties of the probes used to estimate a classical
parameter can be used to attain accuracies that beat the standard
quantum limit.  When qubits are used to construct a quantum probe, it
is known that initializing $n$ qubits in an entangled ``cat state,''
rather than in a separable state, can improve the measurement
uncertainty by a factor of $1/\sqrt{n}$.  We investigate how the
measurement uncertainty is affected when the individual qubits in a
probe are subjected to decoherence.  In the face of such decoherence,
we regard the rate $R$ at which qubits can be generated and the total
duration $\tau$ of a measurement as fixed resources, and we determine
the optimal use of entanglement among the qubits and the resulting
optimal measurement uncertainty as functions of $R$ and $\tau$.
\end{abstract}
\maketitle

\section{Introduction}
\label{sec1}

This paper considers the question of quantum limits on estimating the
value of a parameter that influences the state of a physical
system~\cite{helstrom76a,holevo82a,braunstein94a,braunstein95a,braunstein95b,
braunstein96a,boixo07a,knill07a,giovannetti05a}. We call this system, which is
intrinsically quantum mechanical, a {\em probe}, because it is used to probe the
value of the parameter. The accuracy with which the parameter can be estimated
is determined by the initial quantum state of the probe, the type of interaction
by which the parameter influences the state of the probe, and the readout
measurement that is used to extract information from the probe.

In this paper the probe is always a collection of $n$ qubits.  We
assume that there is sufficient control over the probe qubits to
initialize the probe in any separable or entangled pure state.  We
let $g$ denote the parameter we are estimating.  The effect of $g$
on the $j$th probe qubit is described by the Hamiltonian
\begin{equation}
\label{eq:intro1} H_j(g) = \frac{1}{2} g \sigma_{z;j}\;.
\end{equation}
Here $\sigma_{z;j}$ denotes the Pauli $z$ operator for the $j$th
qubit; similarly, $\sigma_{x;j}$ and $\sigma_{y;j}$  denote the
other two Pauli operators.  The parameter $g$, which has units of
frequency (we choose $\hbar=1$), is a coupling strength.  The
Hamiltonian~(\ref{eq:intro1}) generates a rotation about the $z$
axis of the qubit's Bloch sphere.  The overall influence of the
parameter on the probe is given by the Hamiltonian
\begin{equation}
\label{eq:intro2}
H = \sum_{j=1}^n H_j(g)=\frac{1}{2}g\sum_{j=1}^n\sigma_{z;j}\;.
\end{equation}
The value of $g$ is to be deduced from the change in the state of
the probe.  For simplicity, we assume that the probe qubits do not
have any free Hamiltonian evolution.

Giovannetti, Lloyd, and Maccone~\cite{giovannetti05a} have analyzed a
general scheme of this type.  Their theoretical framework involves
estimating a dimensionless parameter $\varphi$, introduced on the
probe through a unitary transformation $U=e^{-ih\varphi}$, with
$h=\sum_{j=1}^n h_j$.  They do not identify $h$ as a Hamiltonian;
rather it is treated as an arbitrary operator that is the generator
of translations in the variable $\varphi$.  The connection between
our scheme and theirs is established by identifying $h=H/g$
($h_j=H_j/g$) and $\varphi=gT$, where $T$ is the time for which the
Hamiltonian~(\ref{eq:intro2}) acts on the probe.

The chief objective of our analysis, following~\cite{huelga97a}, is
to expand the discussion in~\cite{giovannetti05a} by investigating
how decoherence impacts the accuracy with which the parameter can be
determined.  Thus we assume that in addition to the Hamiltonians
$H_j(g)$, the probe qubits are subject to other influences that can
lead to decoherence. For the decoherence models we consider, the
effects of decoherence manifest themselves at a readily identifiable
rate, which we denote as $\gamma$.  To make the analysis meaningful,
we must impose additional constraints on the probes, since we can
always make decoherence irrelevant by estimating $g$ using a
procedure that is completed in a time much shorter than the time
$\gamma^{-1}$ over which decoherence has a significant effect. Thus
we assume that qubits are made available and initialized into probes
at a rate $R$. What we have in mind is that each probe is assembled
in a time $n/R$ and is then sent immediately through a quantum
channel, where it is subjected to the Hamiltonian~(\ref{eq:intro2})
for a time $T$. If we use $\nu$ probes to estimate $g$, so that the
total number of qubits is
\begin{equation}
N=\nu n\;,
\end{equation}
the total time required is
\begin{equation}
\label{eq:constraint}
\tau=\nu n/R+T\;,
\end{equation}
provided that the quantum channel can accommodate more than one probe
at a time. In our analysis, we assume that the parameter must be
determined in the fixed time $\tau$ ($\tau^{-1}$ can be thought of
roughly as the bandwidth over which a time-varying $g$ is estimated),
that the qubit supply rate $R$ is a fixed resource, and that the
decoherence rate $\gamma$ is a constant. We vary the interaction time
$T$ and the number of qubits in each probe, $n=R(\tau-T)/\nu$, to
achieve the best accuracy in determining~$g$.

A measurement scheme of the sort discussed here appears in slightly
modified forms in several problems of practical importance, such as clock
synchronization~\cite{jozsa00a,chuang00a,preskill00a,revzen03a,burgh05a,
boixo06a}, reference-frame alignment~\cite{bagan01a,chiribella04a},
phase estimation~\cite{gerry03a,dunningham04a,wang05a}, frequency
measurements~\cite{huelga97a,bollinger96a,cappellaro05a}, and position
measurements~\cite{giovannetti02a,giovannetti04a}.

The accuracy with which $g$ can be estimated is closely connected to the
distinguishability of neighboring states of the quantum probe. This
connection is quantified by the generalized uncertainty relations
formulated by Braunstein, Caves, and Milburn~\cite{braunstein96a}.  As
in~\cite{giovannetti05a}, we use these generalized uncertainty relations
to describe the optimal accuracy of parameter estimation.

In Sec.~\ref{sec2}, we review the formalism of generalized
uncertainty relations in the forms suitable for our analysis and
discuss briefly aspects of the assumptions we make about resources
and time scales in our measurement protocol.  Section~\ref{sec3}
reviews the accuracy that can be achieved in the absence of
decoherence, and Sec.~\ref{sec4} investigates how the achievable
accuracies are affected by a general qubit decoherence process.  The
final section provides a short discussion of our results.

\section{Generalized uncertainty relations}\label{sec2}

As a consequence of spending a time $T$ in the quantum channel, the
quantum state of the probe changes from an initial state $\rho_0$ to a
final state
\begin{equation}
\rho(g,T)=
e^{-iHT}{\cal A}_T(\rho_0)e^{iHT}=
e^{-ihgT}{\cal A}_T(\rho_0)e^{ihgT}\;,
\label{eq:rhogT}
\end{equation}
where ${\cal A}_T$ is the superoperator that describes the cumulative
effect of decoherence in the channel.  The final state can always be
written in the form~(\ref{eq:rhogT}), by going to an interaction
picture relative to the Hamiltonian~(\ref{eq:intro2}), but in doing
so, the decoherence superoperator, ${\cal A}_T$, generally becomes
dependent on $g$. In our analysis, however, we assume all decoherence
processes to be invariant under rotations about the $z$ axis (we also
assume that the decoherence is independent and identical from one
qubit to the next), which implies that ${\cal A}_T$ is independent of
$g$ and also means that ${\cal A}_T$ commutes with $e^{-ihgT}$, i.e.,
$\rho(g,T)={\cal A}_T(e^{-ihgT}\rho_0e^{ihgT})$. The final state
contains the information about the value of $g$.  The accuracy with
which we can distinguish the state $\rho(g,T)$ from neighboring
states on the one-parameter path parametrized by $g$ controls the
accuracy in the estimate of $g$.

From the results of measurements on a set of $\nu$ probes, we obtain an
estimate $\gest$ for the value of $g$.  We can quantify the statistical
deviation of the estimate from the true value of $g$ by the
units-corrected deviation from the actual value,
\begin{equation}
\label{eq:gen1}
\delta g \equiv
\left\langle\left(\frac{\gest}{\left|\,d\langle\gest\rangle_g/dg\right|}-g\right)^{\!2\,}\right\rangle^{\!1/2}\;,
\end{equation}
introduced in~\cite{braunstein94a,braunstein96a}.  A lower bound on
$\delta g$ is given by the generalized uncertainty
relation~\cite{helstrom76a,holevo82a,braunstein94a,braunstein95a,braunstein95b,braunstein96a},
\begin{equation}
\label{eq:gen2}
\delta g \geq \frac{1}{\sqrt{\nu}\,(ds/dg)}\;,
\end{equation}
where $ds$ is the ``statistical distance'' between neighboring
quantum states along the trajectory parametrized by $g$.  The
statistical distance is given in terms of the change,
\begin{equation}
d\rho={d\rho(g,T)\over dg}dg\equiv\rho'dg\;,
\end{equation}
in $\rho=\rho(g,T)$ due to a small change $dg$ in the value of $g$:
\begin{equation}
\label{eq:gen3}
\left(\frac{ds}{dg}\right)^{\!2} =
{\mbox{tr}}\Bigl(\rho'{\mathcal{L}}_{\rho}(\rho')\Bigr)\;.
\end{equation}
In the basis $\{|\alpha \rangle\}$ that diagonalizes
$\rho(g,T)=\sum_\alpha p_\alpha|\alpha\rangle\langle\alpha|$, the
superoperator ${\mathcal{L}}_{\rho}$ takes the form
\begin{equation}
\label{eq:gen4}
{\mathcal{L}}_{\rho}(O)
= \sum_{\{\alpha,\,\beta\mid p_{\alpha} + p_{\beta} \neq 0 \}}
\frac{2}{p_{\alpha} + p_{\beta}} O_{\alpha \beta} \,
|\alpha \rangle \langle \beta|\;.
\end{equation}
The operator $\mathcal{L}_\rho(\rho')$ is called the {\em symmetric logarithmic
derivative\/}~\cite{holevo82a} because
\begin{equation}
\rho'={1\over2}\Bigl(\rho\mathcal{L}_\rho(\rho')+\mathcal{L}_\rho(\rho')\rho\Bigr)\;.
\end{equation}
The quantity $(ds/dg)^2={\mbox{tr}}[\rho'{\mathcal{L}}_{\rho}(\rho')]$ is
often called the {\em quantum Fisher information\/}~\cite{holevo82a}.

The generalized uncertainty relations are derived using a quantum
version of the Cramer-Rao
bound~\cite{cramer46a,braunstein94a,braunstein96a}. Generally, this
bound can be achieved only in the case of an optimal measurement on
each probe and, even then, only asymptotically for a large number
$\nu$ of probes, as emphasized by Braunstein~\cite{braunstein06a}.
In our analysis, we explicitly exhibit an optimal measurement, and
we let $\nu_{\rm min}$ denote the number of probes required to
approach the bound within some fixed fractional error.

Consider now the the continuous path in the space of states of the
probe parametrized by $g$.  Nearby points on this path are related
by the derivative
\begin{equation}
\label{eq:gen5a}
\rho'=-iT[h,\rho]=-iT[\hat h,\rho]
=iT\sum_{\alpha,\beta}(p_\alpha-p_\beta)\hat h_{\alpha\beta}\,|\alpha\rangle\langle\beta|\;,
\end{equation}
where here and throughout a hat denotes the difference between a quantity
and its mean value, i.e.,  $\hat h=h-\langle h\rangle$. Plugged into
Eq.~(\ref{eq:gen4}), this gives
\begin{equation}
\label{eq:gen5b}
{\mathcal L}_{\rho}(\rho') =
2iT\sum_{\alpha,\beta} \frac{p_{\alpha} - p_{\beta}}{p_{\alpha} + p_{\beta}}\,
\hat h_{\alpha \beta}\,|\alpha \rangle \langle \beta|
\end{equation}
and
\begin{equation}
\label{eq:gen6}
\left( \frac{ds}{dg} \right)^2=4T^2\Delta^2
\leq 4T^2(\Delta h)^2\;,
\end{equation}
where
\begin{equation}
\label{eq:gen7a}
\Delta^2\equiv{1\over2}
\sum_{\alpha,\beta} \frac{ \left(p_{\alpha} - p_{\beta} \right)^2}{p_{\alpha} + p_{\beta}}
|\hat h_{\alpha \beta}|^{\,2}
\end{equation}
and
\begin{equation}
\label{eq:gen7b}
(\Delta h)^2\equiv\langle\hat h^2\rangle=
{1\over2}\sum_{\alpha,\beta}(p_\alpha+p_\beta)|\hat h_{\alpha\beta}|^2
\end{equation}
is the variance of $h$ with respect to $\rho(g,T)$.  Notice that in
Eqs.~(\ref{eq:gen5b}) and (\ref{eq:gen7a}), we can remove the hat from $h$
without changing anything, whereas in the variance~(\ref{eq:gen7b}), we
cannot do so.  From Eqs.~(\ref{eq:gen2}) and (\ref{eq:gen6}), we obtain
\begin{equation}
\label{eq:gen8}
\delta g \geq
\frac{1}{\sqrt\nu\,2T\Delta}
\geq\frac{1}{\sqrt\nu\,2T\Delta h}\;.
\end{equation}
These are the generalized uncertainty relations in the forms we will
use. Notice that when $\rho(g,T)$ is a pure state, we have
$\Delta=\Delta h$, and thus equality holds in Eq.~(\ref{eq:gen6}) and
in the second inequality in Eq.~(\ref{eq:gen8}).  Notice also that in
the absence of decoherence (i.e., when ${\cal A}_T$ is the unit
superoperator), a pure initial state stays pure, and $\Delta h$ is
independent of~$T$.

The second (weaker) inequality in Eq.~(\ref{eq:gen8}) is generally
easier to work with than the first (stronger) inequality, because
computing the uncertainty $\Delta h$ is usually easier than computing
the corresponding term in the first inequality.  In the case of an
initial pure state in the absence of decoherence, the two
inequalities are equivalent.  In this case, to minimize the
uncertainty in $g$, we should initialize the probe in a state in
which $\Delta h$ is maximal.  We show in Sec.~\ref{sec4} that when
there is decoherence in the probe qubits, which changes an initial
pure state to a mixed state, the weaker inequality is not very
useful, and we are forced to work with the stronger inequality.

For nonGaussian statistics, the quantum Cramer-Rao bound is saturated
only in the limit $\nu \rightarrow \infty$, i.e., when the
measurement process involving an $n$-qubit probe is repeated many
times.   We let $\numin$ denote the minimum number of iterations that
are required for the measurement accuracy to approach the quantum
Cramer-Rao bound within some fixed fractional error.  For the qubit
protocols we analyze, $\numin$ is essentially independent of
protocol, as we discuss further below when we consider measurements
that achieve the bound~(\ref{eq:gen8}); a typical value might be,
say, 50. The need to do at least $\numin$ iterations places a
constraint, $\nu\ge\numin$. Together with the constraint that each
probe must contain at least one qubit, this gives us the following
constraints on $n$:
\begin{equation}
\label{eq:nconstraint}
\label{eq:count1a} 1 \leq n = N/\nu \leq N/\numin\;.
\end{equation}
For these constraints to be consistent, it must be true that
\begin{equation}
R(\tau-T)=N\ge\numin\;.
\end{equation}

As we discussed in the Introduction, we assume in our analysis that the
parameter must be determined in a fixed time $\tau=\nu n/R+T$ and that the
qubit supply rate $R$ is a fixed resource.  For a particular kind of
decoherence, we vary the interaction time $T$ and the number of qubits in
each probe, $n=R(\tau-T)/\nu$, to achieve the best accuracy in
determining~$g$.

A different sort of resource that is dependent on the way the probe
is initialized is the magnitude of $\Delta$.  If we assume that the
energy spread for each of the qubits is fixed, then a way of getting
a large value for $\Delta$ is to initialize the  $n$ qubits in each
probe in an appropriate entangled state.  In a sense, $\Delta$ is
itself a measure of the entanglement or quantum coherence available
for improving the ability to determine $g$.

\section{Measurement accuracy in the absence of decoherence} \label{sec3}

In this section we review the limits on the accuracy of estimating $g$ in
the case where there is no decoherence.  From the generalized uncertainty
relation, we see that the initial state of the probe has a direct bearing
on the optimal accuracy.  We look at two very different initial pure
states of the probe.  In the first case the $n$ probe qubits are
initialized in a product state, and in the second case they are in a
collective entangled state.  For both cases we compute the limit on the
precision with which $g$ can be estimated.  Since there is no decoherence,
the probe state remains pure, and the two inequalities in
Eq.~(\ref{eq:gen8}) are equivalent, because $\Delta=\Delta h$.  Thus in
this section we only need to consider $\Delta h$.  This section also
serves to establish our notation and to summarize the results
in~\cite{giovannetti05a}.

\subsection{Initial pure product state} \label{sec1a}

If the probe is initialized in a pure product state, $\rho_{p}$, of
the $n$ qubits, we have
\begin{equation}
\label{eq:nodec0}
\frac{ds_{\!p}}{dg} =
\sqrt{\sum_{j=1}^n \left( \frac{ds_{\!j}}{dg} \right)^2}\;,
\end{equation}
as shown in  Appendix \ref{appA} (see also~\cite{braunstein96a}).
Here and in the following, the subscript $p$ stands for ``product state.'' The
line element $ds_{\!p}$ is in the space of $n$-qubit density operators, while
$ds_{\!j}$ are line elements in the space of states of each the $n$ probe
qubits.  From Eq.~(\ref{eq:gen8}) we see that the best choice of initial
probe state is one that maximizes
\begin{equation}
\label{eq:nodec1}
\Delta=\Delta h = \sqrt{\sum_{j=1}^n(\Delta h_j)^2}\;,
\end{equation}
where $(\Delta h_j)^2$ is the variance of $h_j$ for the $j$th qubit.  Thus
we have to maximize $\Delta h_j$ for each of the $n$ qubits, and we do so
by initializing each of the $n$ qubits in a pure state lying in the
equatorial plane of the Bloch sphere of states for each qubit.  Here we
choose initial state
\begin{equation}
\label{eq:nodec2}
|\psi_j\rangle = \frac{1}{\sqrt{2}} (|0_j \rangle + |1_j \rangle) \quad {\mbox{or}} \quad
\rho_j = \frac{1}{2} \left( \openone_j + \sigma_{x;\, j} \right)\;.
\end{equation}
The vectors $|0_j \rangle$ and $|1_j \rangle$ denote the eigenstates
of $\sigma_{z;j}$ for the $j$th qubit.  The initial state of the
probe is
\begin{equation}
\label{eq:nodec3}
\rho_{p} = \bigotimes_{j=1}^n \rho_j =
\frac{1}{2^n} \bigotimes_{j=1}^n \left( \openone_j + \sigma_{x;\, j} \right)\;.
\end{equation}

The effect of the coupling to the parameter is to rotate the Bloch vector
of each of the qubits around the $\sigma_z$ axis.  At time $T$, after
passage through the channel, the state of each qubit has rotated through
an angle $gT$, giving a probe state
\begin{equation}
\label{eq:nodec3a}
\rho_{p}(g,T) =
\bigotimes_{j=1}^n \rho_j(g,T) =
\frac{1}{2^n} \bigotimes_{j=1}^n
\left( \openone_j + \sigma_{x;\, j} \cos gT + \sigma_{y;\, j} \sin g T\right)\;.
\end{equation}
We use the evolved state~(\ref{eq:nodec3a}) in our discussion of achieving
the optimal measurement accuracy in Sec.~\ref{nodecachieve}. For the
present, however, since the variance of $h_j$ is unchanged by the
evolution, we can evaluate it using the initial state~(\ref{eq:nodec2}).
This gives a variance $(\Delta h_j)^2=1/4$ for each qubit, and thus
\begin{equation}
\label{eq:nodec6}
(\Delta h)^2 =
n(\Delta h_j)^2 =
\frac{n}{4}\;.
\end{equation}
The generalized-uncertainty bound on the estimate of $g$ becomes
\begin{equation}
\label{eq:nodec7}
\delta g\ge \delta g_{p}=
\frac{1}{T\sqrt{\nu n}}=
 \frac{1}{T\sqrt{N}}=
 \frac{1}{T\sqrt{R(\tau-T)}}\;,
\end{equation}
where $N$ is the total number of qubits used in the measurement scheme,
and $\delta g_p$ is the bound for pure product inputs.

Since the bound depends only on $N$, and not separately on $n$ and
$\nu$, we can always choose $n=1$ without affecting the optimal
measurement accuracy.  This, of course, is the statement that for
product-state inputs, the measurement accuracy is indifferent to
whether we regard the qubits as gathered together into multi-qubit
probes.  To find the optimal bound, all that is left is to adjust the
interaction time $T$ to minimize $\delta g_p$.  Doing so gives
\begin{equation}
\label{eq:minA1}
T = \frac{2}{3}\tau
\end{equation}
and thus
\begin{equation}
\label{eq:minA2}
N = \nu = \frac{1}{3} R\tau\;.
\end{equation}
That there be enough probes to satisfy $\nu\ge\numin$ requires that
${1\over 3}R\tau\ge\numin$.  When ${1\over3}R\tau<\numin$, the
measurement bound is optimized by the choices $n=1$ and $\nu=\numin$.
This gives an interaction time $T=\tau-\numin/R$ that decreases with
$\tau$ until $R\tau=\numin$, at which point it is impossible to
obtain and use $\numin$ probes within the overall duration $\tau$.
This interaction time occurs in every situation we consider, when the
measurement protocol is starved of qubits, so we abbreviate it as
\begin{equation}
\label{eq:Ts}
T_s\equiv\tau-\numin/R\;.
\end{equation}

The optimal bound on measurement accuracy thus takes the form
\begin{equation}
\label{eq:minA3}
\delta g_p=
\cases{
    \displaystyle{{1\over T_s\sqrt{\numin}}}\;,&$1\le R\tau<3\numin$,\cr
    \displaystyle{{3\sqrt3/2\over\tau\sqrt{R\tau}}}\;,&$3\numin\le R\tau$.\cr
      }
\end{equation}
In our resource-based analysis, in which the overall measurement time
$\tau$ and the rate $R$ at which qubits can be supplied are the
resources, the $1/\sqrt{R\tau^3}$ scaling is the signature of the
{\em standard quantum limit}.  The behavior of the bound for $1\le
R\tau<3\numin$ is included for completeness in our subsequent
analysis, but is not so important since it expresses what happens
when the measurement protocol is starved of qubits.

\subsection{Initial pure entangled state} \label{sec1b}

If the probe can be initialized in an entangled state, we can obtain
bigger values of $\Delta h$.  The maximum value is obtained by superposing
two $n$-qubit eigenstates of $h$ corresponding to the lowest and highest
eigenvalues.  Thus we initialize the probe in the ``cat''
state~\cite{giovannetti05a}
\begin{equation}
\label{eq:nodec7g}
|\Psi_c \rangle = \frac{1}{\sqrt{2}} \left( |00\ldots0\rangle + |11\ldots1\rangle \right)\;,
\end{equation}
denoted by the subscript $c$.  The initial density operator $\rho_{c} =
|\Psi_c \rangle \langle \Psi_c|$ can be written in the form
\begin{equation}
\label{eq:nodec8}
\rho_{c} = \frac{1}{2^{n+1}} \left(
\bigotimes_{j=1}^n \left( \openone_j + \sigma_{z;j} \right) +
\bigotimes_{j=1}^n \left( \openone_j - \sigma_{z;j} \right) +
\bigotimes_{j=1}^n \left( \sigma_{x; j} + i\sigma_{y;j} \right) +
\bigotimes_{j=1}^n \left( \sigma_{x; j} -  i\sigma_{y;j} \right)
\right)\;.
\end{equation}

After passage through the quantum channel, the state of the probe becomes
\begin{eqnarray}
\label{eq:nodec8a}
\rho_{c}(g,T)& = &
\frac{1}{2^{n+1}} \left(
\bigotimes_{j=1}^n \left( \openone_j + \sigma_{z;j} \right) +
\bigotimes_{j=1}^n \left( \openone_j - \sigma_{z;j} \right) \right. \nonumber \\
 && \hspace{5 mm} \left. +
 e^{-ingT}\bigotimes_{j=1}^n \left( \sigma_{x; j} + i\sigma_{y;j} \right) +
 e^{ingT}\bigotimes_{j=1}^n \left( \sigma_{x; j} -  i\sigma_{y;j} \right)
 \right)\;.
\end{eqnarray}
We use this form in our discussion of achievability in
Sec.~\ref{nodecachieve}.  Since $\Delta h$ does not change under the
quantum evolution, we can evaluate it using the initial cat state, which
gives
\begin{equation}
\label{eq:nodec10}
(\Delta h)^2 = \frac{n^2}{4}\;.
\end{equation}
The generalized-uncertainty bound becomes
\begin{equation}
\label{eq:nodec12}
\delta g \geq  \delta g_{c} = \frac{1}{T n \sqrt{\nu}} =
\frac{1}{T \sqrt{Nn}} = \frac{\sqrt\nu}{TR(\tau-T)}\;.
\end{equation}
By letting the probe qubits be in a collective entangled state, the
accuracy in our estimate is enhanced by a factor of $1/\sqrt{n}$.

For the cat-state input with no decoherence, it is optimal to make
$\nu$ as small as possible, i.e., $\nu=\numin$.  This puts as many
qubits as possible into each probe consistent with the
constraint~(\ref{eq:nconstraint}), i.e., $n=N/\numin$, which is
clearly optimal in the absence of any decoherence to degrade the
entanglement. To find the optimal bound on the measurement accuracy,
we adjust the interaction time $T$ to minimize $\delta g_c$, giving
\begin{equation}
\label{eq:minA1c}
T = \frac{1}{2}\tau
\end{equation}
and
\begin{equation}
\label{eq:minA2c}
N = n\numin= \frac{1}{2} R\tau\;.
\end{equation}
In order that $\nu\ge\numin$, we require ${1\over 2}R\tau\ge\numin$.
When ${1\over2}R\tau<\numin$, we choose $n=1$ and $\nu=\numin$, which
gives $T=T_s$ and an optimal bound that is the same as for
product-state inputs.

The optimal bound on measurement accuracy thus becomes
\begin{equation}
\label{eq:minA3c}
\delta g_c=
\cases{
    \displaystyle{{1\over T_s\sqrt{\numin}}}\;,&$1\le R\tau<2\numin$,\cr
    \displaystyle{{4\sqrt{\numin}\over R\tau^2}}\;,&$R\tau\ge2\numin$.
      }
\end{equation}
The $1/R\tau^2$ scaling of the cat-state bound is the signature of
the so-called {\em Heisenberg limit\/}; it is to be contrasted with
the corresponding $1/\sqrt{R\tau^3}$ scaling available from product
states. The enhancement available from entanglement is roughly a
factor of $\sqrt{\numin/R\tau}$.  Just as for product states, the
behavior of the bound for $1\le R\tau<2\numin$ is not so important,
as it expresses what happens when the measurement protocol is starved
of qubits.

\subsection{Achieving the optimal measurement accuracy}
\label{nodecachieve}

Estimating $g$ involves making measurements on the probe qubits.  A
strategy that gives an optimal estimate of $g$ is to measure $\sigma_x$ on
all the probe qubits.  This provides a means of estimating $gT$, from
which $g$ can be calculated provided we know the interaction time $T$
accurately.

In the case of an initial pure product state, we can specialize to having
just one qubit in each probe ($n=1$), prepared in the
state~(\ref{eq:nodec2}).  We measure $\sigma_x$ on each qubit.  The
results are averaged over $\nu$ trials to get an accurate estimate for
$g$~\cite{giovannetti05a}.  The expectation value and variance of
$\sigma_x$ with respect to the evolved state in Eq.~(\ref{eq:nodec3a}) are
\begin{equation}
\label{eq:nodec14}
\langle \sigma_x \rangle = \cos g T\;,
\qquad
(\Delta\sigma_x)^2=1-\cos^2\!gT=\sin^2\!gT\;.
\end{equation}
The average of the results over $\nu$ trials, which we denote
$\bar\sigma_x$, has the same expectation value, but its variance
decreases by a statistical factor of $1/\nu$, i.e.,
$\Delta\bar\sigma_x=|\sin gT|/\sqrt\nu$.

We estimate $g$ as $\gest=T^{-1}\arccos\bar\sigma_x$.  When the
uncertainty in $\bar\sigma_x$ is small enough, we can approximate
$\langle\gest\rangle=T^{-1}\arccos\langle\bar\sigma_x\rangle=g$ and
\begin{equation}
\label{eq:nodec18}
\delta g=\Delta\gest=
{\Delta\bar\sigma_x\over|d\langle\bar\sigma_x\rangle/dg|}
= \frac{1}{T \sqrt{\nu}} \;.
\end{equation}
The approximation here is that the datum $\bar\sigma_x$ must be
likely to lie close enough to the expected value
$\langle\bar\sigma_x\rangle$ that a linear approximation to the
arccos function at the operating point is valid. This requires that
$\nu$ be large enough that
$1\gg\Delta\bar\sigma_x=\Delta\sigma_x/\sqrt\nu\sim1/\sqrt\nu$. That
$\nu$ must be large is the expression, in the context of this
particular measurement, of the general fact that the quantum
Cramer-Rao can only be achieved asymptotically; it leads to our
requirement that $\nu\ge\numin\gg 1$.

When the probe is initialized in the cat state, an optimal measurement
strategy is to measure $\sigma_x$ on all $n$ qubits simultaneously and to
multiply all the results together~\cite{giovannetti05a}.  Formally, this
corresponds to measuring
\begin{equation}
\label{eq:nodec19}
\Sigma_x = \bigotimes_{j=1}^n \sigma_{x;j}
\end{equation}
The expectation value and variance of $\Sigma_x$ with respect to the evolved
state~(\ref{eq:nodec8a}) are
\begin{equation}
\label{eq:nodec20}
\langle \Sigma_x \rangle = \cos ng T\;,
\qquad
(\Delta\Sigma_x)^2=1-\cos^2\!ngT=\sin^2\!ngT\;.
\end{equation}
The average of the results over $\nu$ probes has the same expectation
value, but its variance decreases by a statistical factor of $1/\nu$.

We estimate $g$ in the same way as above for product inputs.  The only
difference is the additional factor of $n$ in the rotation angle due to
the coherent rotation of the entangled qubits in each probe.  The
resulting uncertainty in our estimate of $g$ is
\begin{equation}
\label{eq:nodec23}
\delta g = \frac{1}{T n \sqrt{\nu}}\;,
\end{equation}
thus saturating the bound~(\ref{eq:nodec12}).  Notice that the
condition for making a linear approximation to the arccos function is
the same as for product inputs, i.e.,
$1\gg\Delta\bar\Sigma_x=\Delta\Sigma_x/\sqrt\nu\sim1/\sqrt\nu$,
showing that we can take $\numin$ to have the same value for product
and cat-state protocols.

There are technical questions associated with how one resolves the
fringes in Eqs.~(\ref{eq:nodec14}) and (\ref{eq:nodec20}) in order to
zero in on the actual value of $g$.  These questions are well
understood, however, and are irrelevant to our goal of understanding
the effects of decoherence, so we do not consider them further.

\section{Measurement accuracy in the presence of decoherence} \label{sec4}

The previous section reviewed, within the context of our resource-based
analysis, the measurement accuracies that can be obtained in the absence
of decoherence.  In this section we introduce decoherence to see how it
affects the accuracy of parameter estimation.  We consider a general model
for decoherence of the probe qubits, subject to the restrictions that the
decoherence (i)~is independent and identical from one probe qubit to the
next, (ii)~is continuously differentiable and time stationary, and
(iii)~commutes with rotations about the $\sigma_z$ axis.  Since the
interaction Hamiltonian that connects the probe qubits to the parameter
generates rotations about the $\sigma_z$ axis of each of the qubits, the
effect of the third restriction is to separate cleanly the effect of the
parameter from the effects of decoherence.

Decoherence can be described in terms of trace-preserving quantum
operations (completely positive maps) on density operators
\cite{sudarshan61a,stormer63,kraus71a,davies76,breuer02,nielsen00a}. A
quantum operation on single-qubit states is completely specified by the
transformations of the operator basis set consisting of $\openone$,
$\sigma_x$, $\sigma_y$, and $\sigma_z$.  A general time-dependent
trace-preserving map ${\cal A}_t$ on one-qubit states, which commutes with
rotations about the $\sigma_z$ axis, has the form
\begin{eqnarray}
\label{eq:deca0}
{\cal A}_t(\openone) & = & \openone + f(t)\sigma_z\;, \nonumber \\
{\cal A}_t(\sigma_z)& = & g(t)\sigma_z\;, \nonumber \\
{\cal A}_t(\sigma_x \pm i\sigma_y) & = & h_{\pm}(t)(\sigma_x \pm i\sigma_y)\;,
\end{eqnarray}
where $f(t)$, $g(t)$, and $h_+(t)=h_-^*(t)$ are arbitrary functions
of time $t$.  The requirement that the evolution described by ${\cal
A}_t$ be continuously differentiable and time stationary implies that
the derivatives of the quantities on the left of
Eq.~({\ref{eq:deca0}) be linear combinations with constant
coefficients of these same quantities.  Thus we have
\begin{eqnarray}
\label{eq:deca1}
{d{\cal A}_t(\openone)\over dt} & = & \mu\gamma_1{\cal A}_t(\sigma_z)\;, \nonumber \\
{d{\cal A}_t(\sigma_z)\over dt} & = & -\gamma_1{\cal A}_t(\sigma_z)\;, \nonumber
\\
{d{\cal A}_t(\sigma_x \pm i\sigma_y)\over dt} & = & -(\gamma_2 \pm i\omega){\cal
A}_t(\sigma_x \pm i\sigma_y)\;,
\end{eqnarray}
where $\mu$, $\gamma_1$, $\gamma_2$, and $\omega$ are real constants.

The solution of Eqs.~(\ref{eq:deca1}), with ${\cal A}_0={\cal I}$, gives
the most general single-qubit decoherence model that satisfies
restrictions~(ii) and (iii) above:
\begin{eqnarray}
\label{eq:deca2}
{\cal A}_t(\openone) & = & \openone + \mu( 1 - e^{-\gamma_1 t})\sigma_z\;, \nonumber \\
{\cal A}_t(\sigma_z) & = & e^{-\gamma_1 t}\sigma_z\;, \nonumber \\
{\cal A}_t(\sigma_x\pm i\sigma_y) & = & e^{-\gamma_2 t} e^{\mp i\omega t}
(\sigma_x\pm i\sigma_y)\;.
\end{eqnarray}
The dissipation in this model is that of the standard qubit decoherence
model, involving a longitudinal decay time $T_1=1/\gamma_1$ and a
transverse dephasing time $T_2=1/\gamma_2$.  In the limit
$t\rightarrow\infty$, every single-qubit state decays to the state
${1\over2}(\openone + \mu \sigma_z)$, which means that we must have
$-1\le\mu\le1$.    Complete positivity requires that $T_2\le 2T_1$.

In addition to the dissipation, there is a coherent rotation about
the $\sigma_z$ axis by an angle $\omega t$.  If a decoherence process
does introduce such a coherent rotation, it cannot be distinguished
from the rotation produced by the parameter $g$; any procedure for
estimating $g$ would actually estimate $g+\omega$. Throughout the
following, we assume that the decoherence model does not include any
coherent rotation, but for convenience, we incorporate the rotation
due to $g$ into ${\cal A}_t$ by assuming that $\omega=g$ and omitting
the further coherent rotation in Eq.~(\ref{eq:rhogT}).  The mapping
of the Bloch sphere induced by ${\cal A}_t$ (with $\omega=g$) is
illustrated in Fig.~\ref{fig1}.

\begin{figure}[!ht]
\resizebox{16 cm}{6.5 cm}{\includegraphics{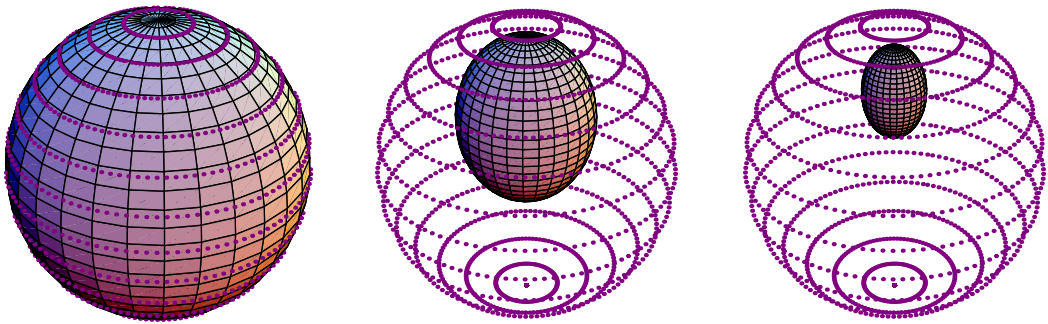}}
\caption{(color online) The transformation of the Bloch sphere under the
map ${\cal A}_t$ of Eq.~(\ref{eq:deca2}) at $\gamma_2 t=0$, $\pi/4$ and
$\pi/2$, with $\mu=3/4$ and $\gamma_1=3\gamma_2/4$. The dotted lines show
the unit sphere.  The coherent rotation due to $\omega=g$ cannot be seen
in these diagrams.}
\label{fig1}
\end{figure}

\subsection{Initial pure product state} \label{decseparable}

We are now prepared to investigate how decoherence affects the theoretical
minimum for $\delta g$ when the probe is initialized in a product state.
At time $t$ after entering the quantum channel, the state of a probe is
given by applying the map~(\ref{eq:deca2}) to the initial
state of each qubit in the pure product state~(\ref{eq:nodec3}):
\begin{equation}
\label{eq:deca3prime}
\rho_{p}(g,t) =
\frac{1}{2^n} \bigotimes_{j=1}^n
\left( \openone_j + \mu(1-e^{-\gamma_1 t})\sigma_{z;j} +
e^{-\gamma_2t}\left( \sigma_{x;j} \cos gt + \sigma_{y;j} \sin gt \right)
\right)\;.
\end{equation}
We first look at the weaker inequality in Eq.~(\ref{eq:gen8}).  At the
time $t=T$ when the probe leaves the quantum channel, we have $\langle h_j
\rangle_T = \frac{1}{2}{\mbox{tr}}[\rho_p(g,T)\sigma_{z;j}]=
\mu(1-e^{-\gamma_1 T})/2$ and $\langle h_j^2\rangle_T=1/4$, giving a
variance
\begin{equation}
\label{deca4b}
(\Delta h)^2
= n\langle\hat h_j^2\rangle_T =
\frac{n}{4}\!\left( 1 - \mu^2\!\left(1-e^{-\gamma_1T} \right)^2 \right)\;.
\end{equation}
The resulting weaker uncertainty-principle bound from
Eq.~(\ref{eq:gen8}) is thus
\begin{equation}
\label{eq:deca4a}
\delta g \geq \delta g_{p}^{(w)}(\gamma_1)
= \frac{1}{T\sqrt{N}}\frac{1}{\sqrt{ 1 - \mu^2 \left(1-e^{-\gamma_1T} \right)^2}}\;.
\end{equation}
Decoherence in the transverse ($\sigma_{x;j}$-$\sigma_{y;j}$) plane
does not appear explicitly in the bound $\delta
g_{p}^{(w)}(\gamma_1)$.  This is because the weaker inequality in
Eq.~(\ref{eq:gen8}) is determined by the variance of $h$, which
depends only on the decoherence along the longitudinal $\sigma_{z;j}$
direction.

Our conclusion is that we should not rely on the weaker inequality in
Eq.~(\ref{eq:gen8}) to provide a good bound on the maximum achievable
measurement accuracy when there is decoherence.  For instance, in the case
where there is only transverse decoherence, i.e., $\gamma_1=0$, the bound
$\delta g_{p}^{(w)}(\gamma_1)$ remains constant at $1/T\sqrt N$, even
though the transverse decoherence ultimately leaves the probe qubits in a
state along the $\sigma_z$ axis where the rotation produced by the
parameter has no effect.  To see the dependence of the measurement
accuracy on the transverse decoherence, we have to use the stronger
inequality in Eq.~(\ref{eq:gen8}).

Turning to that stronger inequality, we need to evaluate $\Delta$ as in
Eq.~(\ref{eq:gen7a}), and for that purpose, we first write the state of
each probe qubit after passage through the channel in diagonal form,
\begin{equation}
\label{eq:dec4b}
\rho_j(g,T)
=\frac{1}{2}\!\left(1+\sqrt{d_1^{\,2}+d_2^{\,2}}\right)|{+}\rangle\langle{+}|
+
\frac{1}{2}\!\left(1-\sqrt{d_1^{\,2}+d_2^{\,2}}\right)|{-}\rangle\langle{-}|
\end{equation}
where
\begin{equation}
\label{eq:dec4c}
d_1 \equiv \mu(1-e^{-\gamma_1T})\;,\quad d_2 \equiv e^{-\gamma_2T}
\end{equation}
and
\begin{eqnarray}
|{+}\rangle&\equiv&
\cos(\theta/2)|0\rangle+e^{igT}\sin(\theta/2)|1\rangle\;,\nonumber \\
|{-}\rangle&\equiv&
\sin(\theta/2)|0\rangle-e^{igT}\cos(\theta/2)|1\rangle\;,
\end{eqnarray}
are the eigenstates of $\rho_j(g,T)$, with
\begin{equation}
\sin\theta={d_2\over\sqrt{d_1^{\,2}+d_2^{\,2}}}\;.
\end{equation}
To evaluate $\Delta_j$ for the $j$th qubit, we need the off-diagonal
matrix element of $h_j={1\over2}\sigma_{z;j}$ in this eigenbasis:
\begin{equation}
(h_j)_{+-}=(h_j)_{-+}^*={1\over2}\langle+|\sigma_{z;j}|-\rangle={1\over2}\sin\theta\;.
\label{eq:dec4e}
\end{equation}

Plugging Eqs.~(\ref{eq:dec4b}) and (\ref{eq:dec4e}) into
Eq.~(\ref{eq:gen7a}), we find that
\begin{equation}
\label{eq:dec4f}
\Delta^2=n\Delta_j^2={n\over4}d_2^{\,2} = {n\over4}e^{-2\gamma_2T}\;,
\end{equation}
from which follows a stronger uncertainty-principle bound,
\begin{equation}
\label{eq:dec4g}
\delta g \ge \delta g_{p}^{(s)}(\gamma_2)
= \frac{e^{\gamma_2 T}}{T \sqrt{N}}=
\frac{e^{ \gamma_2 T}}{T \sqrt{R(\tau-T)}}\;,
\end{equation}
for $\nu$ uses of the quantum probe.  This is a much more reasonable
bound on measurement accuracy, since it depends explicitly on the
transverse decoherence that we expect to make a difference in the
measurement; moreover, the dependence simply degrades the measurement
accuracy exponentially with the number of $T_2$ times for which each
probe is in the quantum channel.

We can show that directly that the bound~(\ref{eq:deca4a}) is weaker
than that of Eq.~(\ref{eq:dec4g}) through the following chain of
inequalities:
\begin{equation}
\sqrt{1-\mu^2(1-e^{-\gamma_1T})^2}\ge
\sqrt{1-(1-e^{-\gamma_1T})^2}=\sqrt{e^{-\gamma_1T}(2-e^{-\gamma_1T})}\ge
e^{-\gamma_1T/2}\ge e^{-\gamma_2T}\;,
\end{equation}
the last of which requires the complete-positivity condition,
$\gamma_2\ge\gamma_1/2$.

Since the probe state is separable at all times, we can choose $n=1$
without affecting the optimal measurement.  What remains is to choose
the interaction time $T$, within the range $0\le T\le\tau$, so as to
minimize the bound $\delta g_p^{(s)}(\gamma_2)$.  There is a single
minimum at $T=T_p$, determined by the equation
$(\gamma_2T_p)^2-(3/2+\gamma_2\tau)\gamma_2T_p+\gamma_2\tau=0$ to
occur at
\begin{equation}
\label{eq:dec4h}
\gamma_2T_p=
{3/2+\gamma_2\tau-\sqrt{(3/2+\gamma_2\tau)^2-4\gamma_2\tau}\over2}\;.
\end{equation}
We cannot use this interaction time when it becomes so short that the
measurement protocol is starved of qubits, i.e., when
$R(\tau-T_p)<\numin$.  In this situation, we choose $n=1$ and
$\nu=\numin$, which gives the interaction time $T_s$ of
Eq.~(\ref{eq:Ts}).

Plugged into Eq.~(\ref{eq:dec4g}), these interaction times give the
optimal value of the bound $\delta g_p^{(s)}(\gamma_2)$, which can be
written in the dimensionless form
\begin{equation}
\label{eq:dec4i}
\sqrt{{R\over\gamma_2}}{\delta g_{p}^{(s)}(\gamma_2)\over\gamma_2}=
\cases{
\displaystyle{\sqrt{{R\over\gamma_2}}\frac{e^{\gamma_2 T_s}}{\gamma_2T_s\sqrt{\numin}}}\;,
&$\displaystyle{{\numin\gamma_2\over R}\le\gamma_2\tau<\gamma_2\Bigl(T_p+{\numin\over R}\Bigr)}$,\cr
\displaystyle{\frac{e^{\gamma_2 T_p}}{\gamma_2T_p\sqrt{\gamma_2(\tau-T_p)}}}\;,
&$\displaystyle{\gamma_2\tau\ge\gamma_2\Bigl(T_p+{\numin\over R}\Bigr)}$.
      }
\end{equation}
We plot the dimensionless optimal interaction time, $\gamma_2T_p$,
and the resulting dimensionless optimal bound~(\ref{eq:dec4i}) as
functions of $\gamma_2\tau$ in Fig.~\ref{fig2}.

\begin{figure}[!ht]
\resizebox{16 cm}{5.5 cm}{\includegraphics{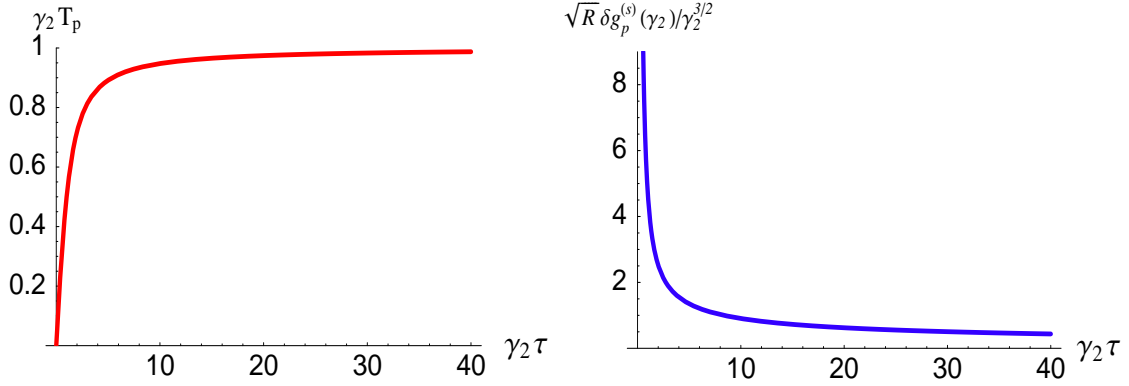}}
\caption{Dimensionless optimal interaction time, $\gamma_2T_p$ of
Eq.~(\ref{eq:dec4h}), and dimensionless optimal
bound~(\ref{eq:dec4i}), plotted as functions of dimensionless
interaction time $\gamma_2 \tau$. In these plots we assume that $R$
is big enough that we do not encounter the situation where the
protocol is starved of qubits, since this situation is of little
interest.} \label{fig2}
\end{figure}

There are two important limits.  When the transverse decoherence has
little effect during the overall time $\tau$, i.e.,
$\gamma_2\tau\ll1$, we find $T_p=2\tau/3$ and an optimal bound that
reduces to the optimal measurement accuracy~(\ref{eq:minA3}) in the
absence of decoherence.  In contrast, for large transverse
decoherence, i.e., $\gamma_2\tau\gg1$, the optimal interaction time
is $T_p=T_2$, and the optimal bound becomes $\delta
g_p^{(s)}(\gamma_2)=e/T_2\sqrt{R\tau}$ or, in terms of the
dimensionless optimal bound, $\sqrt{R/\gamma_2}\delta
g_p^{(s)}/\gamma_2=e/\sqrt{\gamma_2\tau}$. In this case, it is
optimal to have each qubit scoot through the quantum channel in a
dephasing time, before the dephasing can destroy the effect of the
parameter-induced rotation; roughly speaking, each qubit determines
$g$ with accuracy $e/T_2$, which is improved by the statistical
factor $1/\sqrt{R\tau}$ corresponding to the number of qubits used in
time $\tau$.

\subsection{Initial pure entangled state} \label{decentangled}

We now look at the case in which the probe is initialized in an entangled
cat state.  The density operator of the probe, after a time $t$ in the
channel, is obtained by applying the map~(\ref{eq:deca2}) to the initial
cat state~(\ref{eq:nodec8}):
\begin{eqnarray}
\label{eq:deca5}
\rho_{c}(g,t)
&=& \frac{1}{2^{n+1}}
\!\left(
\bigotimes_{j=1}^n \Big( \openone_j + [ e^{-\gamma_1 t} + \mu(1-e^{-\gamma_1t})
] \sigma_{z;j} \Big)+ \bigotimes_{j=1}^n \Big( \openone_j - [ e^{-\gamma_1 t} -
\mu(1-e^{-\gamma_1t}) ] \sigma_{z;j} \Big)\right.\nonumber \\
&&\left.\phantom{ \frac{1}{2^{n+1}}\Biggl[\quad}
+e^{-n\gamma_2 t-ingt} \bigotimes_{j=1}^n ( \sigma_{x; j} + i\sigma_{y;j}) +
e^{-n\gamma_2 t+ ingt} \bigotimes_{j=1}^n ( \sigma_{x; j} - i\sigma_{y;j})
\right)\;.\nonumber \\
\end{eqnarray}

To evaluate the weaker inequality in Eq.~(\ref{eq:gen8}), we need to
evaluate the variance of $h$.  Using Eqs.~(\ref{eq:appb1}), (\ref{eq:appb2}) and
(\ref{eq:appb3}) from Appendix \ref{appB}, we find that at the time $t=T$ when
the probe exits the quantum channel, $\langle h\rangle_T=nd_1/2$ and
\begin{equation}
\label{eq:deca5b}
\langle h^2\rangle_T =
\frac{n}{4} \Big( 1 + (n-1)(e^{-2\gamma_1 T}+d_1^{\,2}) \Big)\;.
\end{equation}
From these we obtain the variance of $h$ at the time $T$ when the probe
exits the quantum channel:
\begin{equation}
\label{eq:deca8}
( \Delta h )^2 = \langle\hat h^2\rangle_T =
\frac{n}{4} \Big( 1 + (n-1)e^{-2\gamma_1 t} - d_1^{\,2} \Big).
\end{equation}
The weaker bound in Eq.~(\ref{eq:gen8}) on the accuracy of estimating $g$ thus
becomes
\begin{equation}
\label{eq:deca11}
\delta g \geq \delta g_{c}^{(w)}(\gamma_1)
=\frac{1}{T \sqrt N}\frac{1}{\sqrt{1 + (n-1)e^{-2 \gamma_1t}-\mu^2 \left( 1 - e^{-\gamma_1t} \right)^2}}\;.
\end{equation}
We do not expect this bound on $\delta g_{c} (\gamma)$ to be particularly
useful because, as for the case of an initial product state, the
decoherence in the transverse directions for the qubits does not come into
the bound at all.

We now look at the bound on measurement accuracy placed by the
stronger inequality in Eq.~(\ref{eq:gen8}). The density operator
$\rho_{c}(g,t)$ is diagonal in the tensor-product basis formed by the
eigenvectors of $\sigma_{z;j}$, except in the two-dimensional
subspace spanned by the vectors
$|00\ldots0\rangle\equiv|\bm{0}\rangle$ and
$|11\ldots1\rangle\equiv|{\bf 1}\rangle$.  We denote this subspace of
the $n$-qubit Hilbert space ${\mathcal H}$ by ${\mathcal K}$.  The
operator $h$ is diagonal in the tensor-product basis formed by
eigenvectors of $\sigma_{z;j}$.  From Eq.~(\ref{eq:gen7a}), we see
that there is no contribution to $\Delta^2$ from the subspace in
which $\rho_{c}(g,t)$ and $h$ are simultaneously diagonal.  Thus, for
computing $\Delta^2$, we can work with the operators
$\bar{\rho}_{c}(g,t)$ and $\bar{h}$ that are obtained by projecting
$\rho_{c}(g,t)$ and $h$ down to the subspace ${\mathcal K}$, i.e.,
\begin{equation}
\label{eq:deca11a}
\bar{\rho}_{c}(g,T)
= \frac{1}{2}\left(
d_{+}|\bm{0}\rangle\langle\bm{0}|+
d_{-}|\bm{1}\rangle\langle\bm{1}|+
d_2^{\,n} e^{-ingT}|\bm{0}\rangle\langle\bm{1}|+
d_2^{\,n} e^{ingT}|\bm{1}\rangle\langle\bm{0}|
\right)
\end{equation}
and
\begin{equation}
\label{eq:deca11b}
\bar{h} = \frac{n}{2}\left(
|\bm{0}\rangle\langle\bm{0}|
-|\bm{1}\rangle\langle\bm{1}|
\right)\;,
\end{equation}
with
\begin{equation}
\label{eq:deca11c}
d_{\pm} \equiv
\left( \frac{1+ e^{-\gamma_1T} \pm d_1}{2} \right)^n +
\left( \frac{1- e^{-\gamma_1T} \pm d_1 }{2} \right)^n\;.
\end{equation}

The next step is to write $\bar{\rho}_{c}(g,T)$ in diagonal form,
\begin{equation}
\label{eq:deca11d}
\bar\rho_c(g,T)
=p_+|\bm{+}\rangle\langle\bm{+}|+p_-|\bm{-}\rangle\langle\bm{-}|\;,
\end{equation}
where
\begin{equation}
\label{eq:deca11e}
p_{\pm}={1\over4}\left(d_+ + d_- \pm \sqrt{(d_+ - d_-)^2+4d_2^{\,2n}}\right)
\end{equation}
are the eigenvalues of $\bar\rho_c$ and
\begin{eqnarray}
|\bm{+}\rangle&\equiv&
\cos(\theta/2)|\bm{0}\rangle+e^{ingT}\sin(\theta/2)|\bm{1}\rangle\;,\nonumber \\
|\bm{-}\rangle&\equiv&
\sin(\theta/2)|\bm{0}\rangle-e^{ingT}\cos(\theta/2)|\bm{1}\rangle\;,
\end{eqnarray}
are the eigenstates, with
\begin{equation}
\sin\theta={2d_2^{\,n}\over\sqrt{(d_+ - d_-)^2+4d_2^{\,2n}}}\;.
\label{eq:deca11f}
\end{equation}

To evaluate $\Delta$, we need the off-diagonal matrix element of $\bar h$ in
this eigenbasis:
\begin{equation}
\bar h_{+-}=\bar h_{-+}^*=\langle\bm{+}|\bar h|\bm{-}\rangle
={n\over2}\sin\theta\;.
\label{eq:deca11g}
\end{equation}
The resulting value of $\Delta^2$ is
\begin{equation}
\label{eq:deca11h}
\Delta^2=(p_+ - p_-)^2|\bar h_{+-}|^2={n^2\over4}e^{-2n\gamma_2T}\;,
\end{equation}
from which follows the stronger uncertainty-principle bound for a
cat-state input,
\begin{equation}
\label{eq:deca11i}
\delta g \ge \delta g_{c}^{(s)}(\gamma_2)
=\frac{e^{n\gamma_2T}}{T n \sqrt{\nu}} =
\frac{e^{n\gamma_2T}}{T \sqrt{nR(\tau-T)}}
= \frac{\sqrt\nu e^{\gamma_2RT(\tau-T)/\nu}}{RT(\tau-T)}
\end{equation}
Aside from being stronger than the bound~(\ref{eq:deca11}), this is a more
sensible bound, since it depends explicitly on the transverse decoherence.
When $\gamma_2=0$, this bound simplifies to the cat-state bound in the
absence of decoherence, Eq.~(\ref{eq:nodec12}).  Moreover, by comparing
with the bound for a product input, Eq.~(\ref{eq:dec4g}), one sees that
this bound retains the $1/\sqrt n$ advantage purchased by using an
entangled input, but at the price of a decoherence rate that is $n$ times
faster.

We have assumed that both $n$ and $T$ are controllable parameters in
the estimation scheme we are considering, with $\nu$ determined by
Eq.~(\ref{eq:constraint}).  To minimize the bound~(\ref{eq:deca11}),
we use the second form, from which $\nu$ has been eliminated. The
values for $n$ and $T$ that minimize $\delta g_{c}^{(s)}$ can then be
found by solving simultaneously the two equations,
\begin{eqnarray}
\label{eq:catdec1}
&&0=\frac{\partial\delta g_{c}^{(s)}}{\partial n}=
\frac{e^{n\gamma_2 T}}{2 n T \sqrt{nR(\tau -T)} }(2n \gamma_2 T -1)\;,
\nonumber \\
&&0=\frac{\partial\delta g_{c}^{(s)}}{\partial T}=
\frac{e^{n\gamma_2 T}}{2 T^2(\tau-T) \sqrt{nR(\tau -T)} }[3T+2n\gamma_2 T(\tau-T)-2\tau]\;,
\end{eqnarray}
which give $n=1/\gamma_2\tau$ and $T=\tau/2$.  The determinant and
trace of the Hessian of $\delta g_{c}^{(s)}$, with respect to $n$ and
$T$, evaluated at this point, are both positive, showing that it is
indeed a minimum.  The minimum value of the bound is
\begin{equation}
\label{eq:catdec2} \delta g_{c}^{(s)}(\gamma_2)
=\frac{2\sqrt{2e}}{\tau \sqrt{R/\gamma_2}}\;.
\end{equation}

This minimum cannot always be attained, however, because we have the
additional constraints of Eq.~(\ref{eq:nconstraint}), i.e., $1\leq n
\leq N/\numin=R(\tau-T)/\numin=R\tau/2\numin$, which do not always
allow us to choose $n$ equal to the optimal value $1/\gamma_2\tau$.
If $\gamma_2 \tau$ does not lie between $2\numin/R\tau$ and 1, we
have to choose a value for $n$ that lies on the boundary of allowed
values.  There are two cases to consider. If the decoherence rate is
high, i.e., $\gamma_2\tau\ge1$, we choose $n=1$, thus using probes
consisting of individual qubits to estimate $g$, in which case the
analysis reduces to that of the preceding subsection.  Notice that
$\gamma_2\tau=1$ gives $\gamma_2(T_p+\numin/R)=1/2+\numin\gamma_2/R$.
Thus if $2\numin\gamma_2/R\le1$, the second case in
Eq.~(\ref{eq:dec4i}) applies whenever $\gamma_2\tau\ge1$.  If,
however, $2\numin\gamma_2/R>1$, the protocol begins to be starved
of qubits for some $\gamma_2\tau>1$, and there is no situation
where cat states offer any advantage.  Throughout the following,
therefore, we assume that $2\numin\gamma_2/R\le1$.

If the decoherence is small, i.e.,
\begin{equation}
\label{eq:catdec3a}
\gamma_2\tau < 2\numin/R\tau
\quad\Longleftrightarrow\quad
\tau^2 < 2\numin/\gamma_2R\;,
\end{equation}
we use the largest cat state that can be constructed from the
available resources, thus choosing $\nu=\numin$.  Using the last form
in Eq.~(\ref{eq:deca11i}), with $\nu=\numin$, we find that $\delta
g_c^{(s)}$ has extrema for $T(\tau-T)=\numin/\gamma_2R$ and
$T=\tau/2$.  We discard the first possibility because it is
inconsistent with the constraint~(\ref{eq:catdec3a}), i.e.,
$\numin/\gamma_2R=T(\tau-T)\le\tau^2/4<\numin/2\gamma_2R$. Moreover,
the second derivative of $\delta g_c^{(s)}$ with respect to $T$,
evaluated at $T=\tau/2$, is strictly positive when
Eq.~(\ref{eq:catdec3a}) is satisfied, showing that $T=\tau/2$ gives a
minimum.  The optimal interaction time is again $T=\tau/2$
($n=R\tau/2\numin$), and the minimum value of the bound becomes
\begin{equation}
\label{eq:catdec4}
\delta g_{c}^{(s)}(\gamma_2) = \frac{4\sqrt{\numin}}{R\tau^2}
e^{ \gamma_2 R\tau^2/ 4\numin}\;.
\end{equation}
Notice that Eq.~(\ref{eq:catdec4}) reduces to the second case in
Eq.~(\ref{eq:minA3c}) when there is no decoherence, i.e., when
$\gamma_2 = 0$.

The first case in Eq.~(\ref{eq:minA3c}) reminds us that one further
case occurs at very short times, when the protocol is starved of
qubits.  In particular, when $R\tau/2\numin<1$, we must choose $n=1$
and $\nu=\numin$, leading to the familiar interaction time $T_s$
of Eq.~(\ref{eq:Ts}).

\begin{table}
\caption{Minimum value of bound on estimating $g$ using cat-state probes
with available resources deployed optimally.  The table assumes that
$2\numin\gamma_2/R\le1$. \label{table}}
\begin{ruledtabular}
\begin{tabular}{cccccc}
Range of $\gamma_2\tau$                                                                     &$T$                            &$n$                &$\nu$                  &$N$            &$\delta g_c^{(s)}(\gamma_2)$                                               \\\vspace{6pt}
$\displaystyle{{\numin\gamma_2\over R}\le\gamma_2\tau<{2\numin\gamma_2\over R}}$            &$T_s$ of Eq.~(\ref{eq:Ts})     &1                  &$\numin$               &$\numin$       &$\displaystyle{{e^{\gamma_2T_s}\over T_s\sqrt{\numin}}}$                   \\\vspace{6pt}
$\displaystyle{{2\numin\gamma_2\over R}\le\gamma_2\tau<\sqrt{{2\numin\gamma_2\over R}}}$    &$\tau/2$                       &$R\tau/2\numin$    &$\numin$               &$R\tau/2$      &$\displaystyle{{4\sqrt\numin\over R\tau^2}e^{\gamma_2 R\tau^2/4\numin}}$   \\\vspace{6pt}
$\displaystyle{\sqrt{{2\numin\gamma_2\over R}}\le\gamma_2\tau<1}$                           &$\tau/2$                       &$1/\gamma_2\tau$   &$\gamma_2 R\tau^2/2$   &$R\tau/2$      &$\displaystyle{{2\sqrt{2e}\over\tau\sqrt{R/\gamma_2}}}$                    \\\vspace{6pt}
$\gamma_2\tau\ge1$                                                                          &$T_p$ of Eq.~(\ref{eq:dec4h})  &1                  &$R(\tau-T_p)$          &$R(\tau-T_p)$  &$\displaystyle{{e^{\gamma_2 T_p}\over T_p\sqrt{R(\tau-T_p)}}}$
\end{tabular}
\end{ruledtabular}
\end{table}

\begin{figure}[!ht]
\resizebox{12 cm}{7.5 cm}{\includegraphics{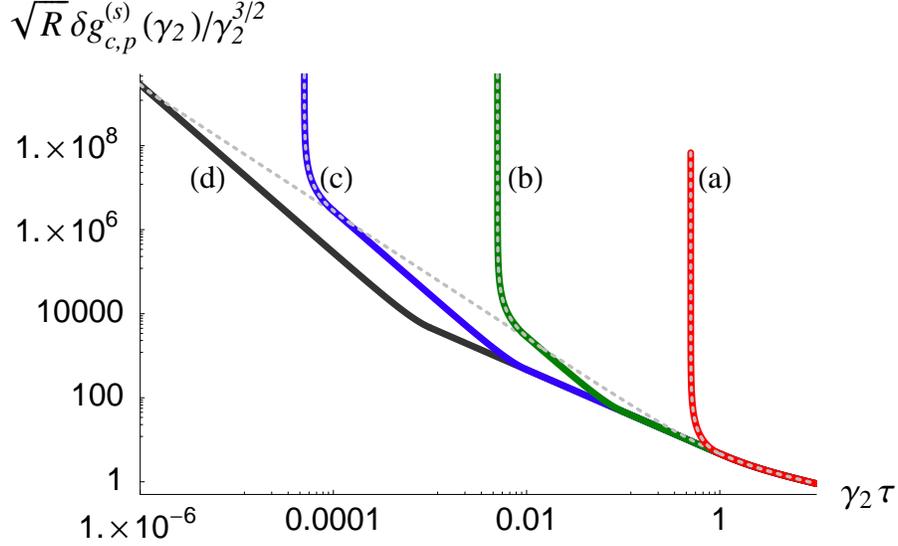}} \caption{(color
online) The four thick lines labeled (a), (b), (c), and (d) show the
dimensionless optimal bound~(\ref{eq:thebound}) for cat-state inputs,
plotted as a function of dimensionless interaction time
$\gamma_2\tau$, for $\numin=50$ and $\sqrt{R/2\numin\gamma_2}=1$, 10,
100, and $1\,000$, i.e., $\sqrt{R/\gamma_2}=10$, 100, $1\,000$, and
$10\,000$, respectively. Note that both axes use a logarithmic scale.
The use of cat states provides no advantage for $\gamma_2\tau\ge 1$.
The two regions where cat states provide an advantage (second and
third rows of Table~\ref{table}) are absent for
$\sqrt{R/2\numin\gamma_2}=1$, but become apparent for the other three
values of $\sqrt{R/2\numin\gamma_2}$. In terms of the dimensionless
bound, the transition region between high and low decoherence (third
row in Table~\ref{table}) has a form independent of $R/\gamma_2$, but
extends to smaller values of $\gamma_2\tau$ as $R/\gamma_2$
increases.  For small enough $\gamma_2\tau$, the protocol is starved
of qubits, and cat states again provide no advantage over product
states.  The thin dotted lines show the dimensionless product-state
bound~(\ref{eq:dec4i}) for the same four values of
$\sqrt{R/2\numin\gamma_2}$.  The product-state bound agrees with the
cat-state bound in the high-decoherence region ($\gamma_2\tau\ge1$)
and with the corresponding cat-state bound in the region where the
protocol is starved of qubits
($\numin\gamma_2/R\le\gamma_2\tau<2\numin\gamma_2/R$); in between,
where cat states provide an advantage, the product-state bound is
independent of $R/\gamma_2$.} \label{fig3}
\end{figure}

We can now piece together the various regions that govern the optimal
bound on the estimate of $g$ using cat-state probes with the
available resources deployed in the optimal fashion.  The results are
summarized in Table~\ref{table}.  The top row is the case where the
protocol is starved of qubits; the second row is the case of low
decoherence, for which the probes are prepared in cat states
containing as many qubits as allowed by the need to have at least
$\numin$ probes; the bottom row is the case of high decoherence, for
which the probes are individual qubits; and the middle row describes
the transition from high decoherence to low decoherence.  The two
regions where cat states play a role exist when
$2\numin\gamma_2/R<1$.  The dimensionless bound introduced in
Eq.~(\ref{eq:dec4i}) is given by
\begin{equation}
\sqrt{R\over\gamma_2}{\delta g_c^{(s)}(\gamma_2)\over\gamma_2}=
\cases{
\displaystyle{\sqrt{{R\over\gamma_2}}\frac{e^{\gamma_2 T_s}}{\gamma_2T_s\sqrt{\numin}}}\;,
&$\displaystyle{{\numin\gamma_2\over R}\le\gamma_2\tau<{2\numin\gamma_2\over R}}$,\cr
\displaystyle{\sqrt{{\gamma_2\over R}}{4\sqrt\numin\over(\gamma_2\tau)^2}e^{(R/\gamma_2)(\gamma_2\tau)^2/4\numin}}\;,
&$\displaystyle{{2\numin\gamma_2\over R}\le\gamma_2\tau<\sqrt{{2\numin\gamma_2\over R}}}$,\cr
\displaystyle{{2\sqrt{2e}\over\gamma_2\tau}}\;,
&$\displaystyle{\sqrt{{2\numin\gamma_2\over R}}\le\gamma_2\tau<1}$,\cr
\displaystyle{{e^{\gamma_2 T_p}\over\gamma_2 T_p\sqrt{\gamma_2(\tau-T_p)}}}\;,
&$\gamma_2\tau\ge1$.
}
\label{eq:thebound}
\end{equation}
This dimensionless optimal bound is plotted in Fig.~\ref{fig3} for
the choice $\numin=50$ and for several values of $\sqrt{R/\gamma_2}$.

The conclusion to be reached from the results summarized in
Table~\ref{table} and Fig.~\ref{fig3} is that cat-state entanglement
is only useful for improving the estimate of $g$ when one wants to
estimate $g$ on a time scale that is shorter than the time scale over
which decoherence acts.

\subsection{Achieving the optimal measurement accuracy}\label{decachieve}

We now examine the effectiveness of the measurement strategies
described in Sec.~\ref{nodecachieve} in the presence of decoherence
in the probe qubits.  When the input state is a product state, we can
again specialize to the case where there is only one qubit in each
probe ($n=1$) and $\nu = N$.  We measure $\sigma_x$ on each qubit and
average over the results of $\nu$ trials to obtain the estimate of
$g$.  After an interaction time $t=T$, the expectation value and
variance of $\sigma_x$ for a single qubit in the evolved
state~(\ref{eq:deca3prime}) are
\begin{equation}
\label{eq:deca28}
\langle\sigma_x \rangle = e^{-\gamma_2 T} \cos g T\;,
\quad
(\Delta\sigma_x)^2 = 1-e^{-2\gamma_2 T}\cos^2\!gT\;.
\end{equation}

Averaging over the results of $\nu$ trials yields the quantity
$\bar{\sigma}_x$, which has the same expectation value as $\sigma_x$,
but has variance reduced to
$(\Delta\bar\sigma_x)^2=(\Delta\sigma_x)^2/\nu$, and we estimate $g$
as $g=\gest=T^{-1}\arccos(e^{\gamma_2T}\bar\sigma_x)$.  After many
trials, the variance of $\bar\sigma_x$ becomes small enough that we
can approximate
$\langle\gest\rangle=T^{-1}\arccos(e^{\gamma_2T}\langle\bar\sigma_x\rangle)=g$
and
\begin{equation}
\label{eq:deca32}
\delta g = \Delta \gest =
{\Delta\bar\sigma_x\over|d\langle\bar\sigma_x\rangle/dg|} =
\frac{e^{\gamma_2T}}{T\sqrt{\nu}}
\frac{\sqrt{1-e^{-2\gamma_2T}\cos^2\!g T}}{|\sin g T|}\;.
\end{equation}
If we use this straightforward method of estimating $g$ by averaging
measurements of $\sigma_x$ on all qubits, achieving the
bound~(\ref{eq:dec4g}) for determining $g$, with the interaction time
adjusted to the optimal value $T=T_p$, requires $|\sin gT_p|=1$. Even
though $g$ is not known, this can be accomplished by using feedback
onto the rotation of the qubits to find this sweet spot on the fringe
pattern.  An alternative to feeding back onto the rotation of the
qubits is feedback to rotate the quantity measured in the equatorial
plane of the Bloch sphere until the desired operating point is
achieved.

When the probe is initialized in a cat state, we have seen that the
optimal strategy, in the absence of decoherence, is to measure
$\sigma_x$ on all $n$ qubits simultaneously.  The same measurement
works when there is decoherence.  Using Eqs.~(\ref{eq:nodec19}) and
(\ref{eq:deca5}), we find that after an interaction time $t=T$,
\begin{equation}
\label{eq:deca35}
\langle \Sigma_x \rangle = e^{-n \gamma_2 T} \cos n g T\;,
\quad
(\Delta\Sigma_x )^2 = 1-e^{-2n\gamma_2 \tau} \cos^2\!n g T\;.
\end{equation}
We can convert the results for product-state inputs to this case by
the substitution $T\rightarrow nT$, so the average over $\nu$ probes
leads to
\begin{equation}
\label{eq:deca38}
\delta g =
\frac{e^{n\gamma_2T}}{Tn\sqrt{\nu}} \frac{\sqrt{1-e^{-2 n\gamma_2T}\cos^2\!ngT }}{|\sin ngT|}\;.
\end{equation}
We can saturate the bound~(\ref{eq:deca11i}) on estimating $g$, for the
appropriate interaction time $T$, by using feedback to operate at a
point where $|\sin ngT|=1$.

One last point concerns the question of making a linear approximation
to the arccos function, which allows us to relate the mean and
variance of $\gest$ directly to the mean and variance of
$\bar\Sigma_x$.  This approximation requires that $\nu$ be large
enough that $1\gg
e^{n\gamma_2T}\Delta\bar\Sigma_x=e^{n\gamma_2T}\Delta\Sigma_x/\sqrt\nu$.
The results summarized in Table~\ref{table} show that for the optimal
choices of $n$ and $T$, it is always true that $n\gamma_2T\sim1$,
showing that the requirement is that $\nu$ be large in a way that is
independent of the details of the protocol.

\section{Discussion} \label{Conclusion}

In this paper we have studied quantum limits on determining a
frequency $g$ that controls the rate at which qubits rotate about the
$z$ axis of the Bloch sphere.  The question of determining $g$ is the
same as the problem of distinguishing qubit states that differ by
having been subjected to different rotations.  The quantum limits on
determining $g$ are well known~\cite{giovannetti05a}: if $n$ qubits
are prepared in product states, the uncertainty in determining $g$
scales as $1/\sqrt n$, the standard quantum limit or shot-noise
limit, whereas if the same $n$ qubits are prepared in an entangled
cat state, the uncertainty scales as $1/n$, which is called the
Heisenberg limit.  Our purpose in this paper has been to investigate,
following~\cite{huelga97a}, how these scalings change when the qubits
are subjected to independent, but identical decoherence processes
during the measurement.

The decoherence model we consider is the most general continuous-time
process that is invariant under rotations about the $z$ axis.  This
results in a standard qubit decoherence model in which qubits decay
with a time constant $T_1$ and lose phase coherence with a time
constant $T_2$.  It is the phase-coherence time $T_2$ that is
important for efforts to estimate $g$; cat-state entanglement is
only useful for times of order or smaller than $T_2$.

To make the analysis meaningful, we introduce as resources the rate
$R$ at which qubits are supplied and the overall time $\tau$ that one
has available for estimating $g$.  If one does not introduce these
resources, one can achieve any desired accuracy in estimating $g$ by
assuming that one can assemble an arbitrarily large number of qubits
in a cat state in a time much shorter than $T_2$ or, more easily, by
assuming that one can take as long as desired to determine $g$ using
an arbitrarily large number of qubits prepared in a product state.
Once decoherence becomes a consideration, the uncertainty in
estimating $g$ should be written in terms of the decoherence time
$T_2$ and the relevant resources, $R$ and $\tau$, not directly in
terms of the number of qubits used.

The results of our analysis, summarized in Table~\ref{table} and
Fig.~\ref{fig3}, show that cat-state entanglement is useless if
$\tau\ge T_2$.  When $T_2$ is much larger than $\tau$, the results
show that one should put as many qubits as possible in each cat-state
probe.  For intermediate decoherence, it is not that cat-state
entanglement is useless, but rather that one should make a judicious,
optimal choice of how many qubits to include in each cat-state probe.
The overall conclusion is that entanglement is only useful when one
can make the effects of decoherence small on the time scale over
which one must estimate $g$. While this conclusion is reached here
for a special model of measurements on qubits, it is generally true
for quantum-limited measurements in the face of decoherence.

Our analysis highlights one further point, having to do with using
the quantum Cramer-Rao bound to determine quantum limits.  As long as
one is interested only in measurements involving pure states, the
form of the Cramer-Rao bound as a generalized uncertainty principle
is sufficient for investigating bounds on measurement accuracy.  Once
decoherence is introduced, however, inevitably leading to
measurements on mixed states, one must use the stronger form of the
Cramer-Rao bound involving the Fisher information to obtain
meaningful bounds on measurement accuracy.

\acknowledgments This work was supported in part by U.S.~Office of
Naval Research Contract No.~N00014-07-1-0304.  We thank S.~Boixo and
A.~Datta for useful suggestions and discussions.

\appendix

\section{Generalized uncertainty relations for product states}\label{appA}

Consider a continuous trajectory in the space of product states of
$n$ qubits parametrized by $X$,
\begin{equation}
\label{eq:appa1}
\rho_{p}(X) = \bigotimes_{j=1}^{n} \rho_j(X)\;,
\end{equation}
If the value of $X$ changes by a small amount $dX$, the change in $\rho_j$
can be written as
\begin{equation}
\label{eq:appa2}
\rho_j \rightarrow \rho_j + dX \rho_j'
= \sum_{\alpha_j} dp_{\alpha_j} |\alpha_j \rangle \langle \alpha_j| + e^{-i
h_jdX} \rho_j   e^{i h_jdX}\;.
\end{equation}
The vectors $\{|\alpha_j \rangle \}$ make up the eigenbasis of $\rho_j$, with
eigenvalues $p_{\alpha_j}$, i.e.,
\begin{equation}
\label{eq:appa3}
\rho_j = \sum_{\alpha_j} p_{\alpha_j} |\alpha_j \rangle \langle \alpha_j|\;.
\end{equation}
The operators $h_j$ are the generators of translations in $X$ for each of
the $n$ systems, while $dp_{\alpha_j}$ are small changes in the
eigenvalues of $\rho_j$ due to the small change in $X$.  Notice that in
the presentation of Sec.~\ref{sec2}, the fact that ${\cal A}_T$ is
independent of the parameter $g$ means that the eigenvalues of $\rho(g,T)$
do not change with $g$; thus the terms having to do with eigenvalue
changes do not appear in that discussion.

Keeping terms to linear order in $dX$, we have
\begin{equation}
\label{eq:appa5}
\rho_j' =
\sum_{\alpha_j} \frac{dp_{\alpha_j}}{dX} |\alpha_j \rangle \langle \alpha_j| -
i[h_j,\rho_j]\;.
\end{equation}
The corresponding change in the overall state,
\begin{equation}
\label{eq:appa6}
\rho_{p} \rightarrow \rho_{p} + dX \rho_{p}' = \rho_{p} + dX \sum_{j=1}^n
\rho_j' \bigotimes_{k \neq j} \rho_{k}\;,
\end{equation}
gives us
\begin{equation}
\label{eq:appa7}
\rho_{p}'=
\sum_{j=1}^n \left[ \sum_{\alpha_j} \frac{dp_{\alpha_j}}{dX} |\alpha_j \rangle
\langle \alpha_j|
- i[h_j,\rho_j]\right] \bigotimes_{k \neq j} \rho_k\;.
\end{equation}

Our objective is to obtain an expression for a line element $ds_{\!p}^2$ in
the space of density operators $\rho_{p}$ that measures the
distinguishability of neighboring quantum states. Following
Eq.~(\ref{eq:gen3}), we have
\begin{equation}
\label{eq:appa8}
\left( \frac{ds_{\!p}}{dX} \right)^2 ={\mbox{tr}}\!\Bigl[ \rho_{p}'
{\mathcal{L}}_{\rho_{p}} (\rho_{p}') \Bigr]\;.
\end{equation}

We start by computing ${\mathcal{L}}_{\rho_{p}} (\rho_{p}')$ using the
definition in Eq.~(\ref{eq:gen4}). Noting that $\rho_{p}$ is diagonal in
the tensor product basis furnished by $\{|\alpha_j \rangle\}$ for each of
the systems and using Eq.~(\ref{eq:appa7}) we obtain
\begin{eqnarray}
\label{eq:appa11}
{\mathcal{L}}_{\rho_{p}} (\rho_{p}')
& = & \sum_{\beta_1,\beta_2 \ldots , \beta_n}
\sum_{\delta_1,\delta_2 \ldots ,\delta_n}
\frac{2}{p_{\beta_1}\cdots p_{\beta_n} + p_{\delta_1} \cdots p_{\delta_n}}
\nonumber \\
&& \quad\times \left \langle \beta_1 \ldots \beta_n \left|\,\sum_{j=1}^n
\left[ \sum_{\alpha_j} \frac{dp_{\alpha_j}}{dX} |\alpha_j \rangle \langle
\alpha_j| - i[h_j,\rho_j] \right]
\bigotimes_{k \neq j} \rho_k \right| \delta_1 \ldots \delta_n \right\rangle
|\beta_1 \ldots \beta_n \rangle \langle \delta_1 \ldots
\delta_n|\;.\nonumber\\
\end{eqnarray}
Simplifying this expression gives
\begin{eqnarray}
\label{eq:appa13}
{\mathcal{L}}_{\rho_{p}} (\rho_{p}') & = &
\sum_{j=1}^n \left[ \sum_{\alpha_j}
\frac{(dp_{\alpha_j}/dX)}{p_{\alpha_j}}|\alpha_j \rangle \langle \alpha_j |
+ 2i\sum_{\alpha_j, \beta_j} \frac{p_{\alpha_j} - p_{\beta_j}}{p_{\alpha_j} +
p_{\beta_j}} [h_j]_{\alpha_j\beta_j}
|\alpha_j \rangle \langle \beta_j|  \right]
\bigotimes_{k \neq j} \openone_k \nonumber \\
& = & \sum_{\alpha=1}^n {\mathcal L}_{\rho_j} (\rho_j') \bigotimes_{k \neq j}
\openone_k\;,
\end{eqnarray}
which leads to
\begin{equation}
\label{appa14}
\left(\frac{ds_{\!p}}{dX} \right)^2 =
\sum_{j=1}^n {\mbox{tr}} \!\Bigl[ \rho_j'{\mathcal L}_{\rho_j} (\rho_j')
\Bigr] \times \prod_{k \neq j} {\mbox{tr}} \rho_k=
\sum_{j=1}^n {\mbox{tr}}\!\Bigl[ \rho_j' {\mathcal L}_{\rho_j} (\rho_j')
\Bigr]
=\sum_{j=1}^n\left(\frac{ds_{\!j}}{dX}\right)^2 \;,
\end{equation}
where we use ${\mbox{tr}} \rho_j' = 0$.  In the special case where all
$\rho_j$ are identical and equal to $\rho$ and when changes in $X$ affect
all the systems in the same way, we have
\begin{equation}
\label{eq:appa15}
\frac{ds_{\!p}}{dX}
= \sqrt{n} \frac{ds}{dX}  \quad {\mbox{where}} \quad \left( \frac{ds}{dX}
\right)^2
={\mbox{tr}} \Bigl[ \rho' {\mathcal L}_{\rho} (\rho') \Bigr]\;.
\end{equation}

\section{Useful identities involving Pauli operators}\label{appB}

A few identities involving products of Pauli operators that are used in our
calculations are listed below:
 \begin{eqnarray}
\label{eq:appb1}
\left( \sum_{j=1}^n \sigma_{z;j} \bigotimes_{j \neq k} \openone_k \right)
\bigotimes_{j=1}^n \left( \openone_j \pm \sigma_{z;j} \right) & = &
\pm n\bigotimes_{j=1}^n \left( \openone_j \pm \sigma_{z;j} \right)\;,
\nonumber \\
\left( \sum_{j=1}^n \sigma_{z;j} \bigotimes_{j \neq k} \openone_k \right)
\bigotimes_{j=1}^n \left( \sigma_{x;j} \pm i\sigma_{y;j} \right) & = &
\pm n\bigotimes_{j=1}^n \left( \sigma_{x;j} \pm i\sigma_{y;j} \right)\;,
\nonumber \\
\left(  \sum_{j\neq k =1}^n \sigma_{z;j} \otimes \sigma_{z; \, k}
\bigotimes_{l \neq j,k} \openone_l \right)
\bigotimes_{j=1}^n \left( \openone_j \pm \sigma_{z;j} \right) & = &
(n^2-n) \bigotimes_{j=1}^n \left( \openone_j \pm \sigma_{z;j} \right)\;,
\nonumber \\
\left(  \sum_{j\neq k =1}^n \sigma_{z;j} \otimes \sigma_{z; \, k}
\bigotimes_{l \neq j,k} \openone_l \right)
\bigotimes_{j=1}^n \left( \sigma_{x;j} \pm i\sigma_{y;j} \right) & = &
(n^2 - n) \bigotimes_{j=1}^n \left( \sigma_{x;j} \pm i\sigma_{y;j}
\right)\;,
\end{eqnarray}
\begin{eqnarray}
\label{eq:appb2}
\left( \sum_{j=1}^n \sigma_{z;j} \bigotimes_{j \neq k} \openone_k \right)
\bigotimes_{j=1}^n \left( \openone_j \pm A \sigma_{z;j} \right) & = &
\pm A \sum_{j=1}^n \left(\openone_j \pm A^{-1} \sigma_{z;j} \right)
\bigotimes_{j\neq k=1}^n \left( \openone_k \pm A\sigma_{z; \, k} \right)\;,
\end{eqnarray}
\begin{eqnarray}
\label{eq:appb3}
\left( \sum_{j\neq k =1}^n \sigma_{z;j} \otimes \sigma_{z; \, k} \bigotimes_{l
\neq j,k} \openone_l \right)
\bigotimes_{j=1}^n \left( \openone_j \pm A \sigma_{z;j} \right) & = &
A^2 \sum_{j\neq k=1}^n \left(\openone_j \pm A^{-1} \sigma_{z;j} \right)
\otimes \left(\openone_k \pm A^{-1} \sigma_{z; \, k} \right) \nonumber \\
&& \hspace{2.5 cm} \bigotimes_{l \neq j,k} \left( \openone_l \pm A \sigma_{z;
\, l} \right)\;.
\end{eqnarray}

\bibliography{measure}

\end{document}